# Hydrogen Isotope Trapping in Al-Cu Binary Alloys

Paul Chao [a,b] and Richard A. Karnesky [a,1]

[a] Sandia National Laboratories, Livermore, CA, USA

[b] Carnegie Mellon University, Pittsburgh, PA, USA

## Abstract

The trapping mechanisms for hydrogen isotopes in Al-$X$ Cu (0.0 at. % < $X$ < 3.5 at. %) alloys were investigated using thermal desorption spectroscopy (TDS), electrical conductivity, and differential scanning calorimetry. Constant heating rate TDS was used to determine microstructural trap energies and occupancies. In addition to the trapping states in pure Al reported in the literature (interstitial lattice sites, dislocations, and vacancies), a trap site due to Al-Cu intermetallic precipitates is observed. The binding energy of this precipitate trap is (18 ± 3) kJ·mol$^{-1}$ (0.19 ± 0.03 eV). Typical occupancy of this trap is high; for Al-2.6 at. % Cu (a Cu composition comparable to that in AA2219) charged at 200 °C with 130 MPa D$_2$ for 68 days, there is ca. there is 3.15x10$^{-7}$ mol D bound to the precipitate trap per mol of Al, accounting for a third of the D in the charged sample.

---

[1] Corresponding author.
E-mail address: rakarne@sandia.gov (R.A. Karnesky)

# 1. Introduction

Current materials used for hydrogen storage such as steels are prone to hydrogen embrittlement, in part due to hydrogen trapping at defects. This embrittlement causes concerns over structural integrity and safety. While austenitic stainless steels (SS) are less prone to these issues than high-strength ferritic steels, embrittlement is still a concern at higher pressures and temperatures [1]. As interest in hydrogen as an alternative fuel source has risen, there has been a corresponding search for other materials for hydrogen transport and storage. Aluminum alloys are promising materials for many structural applications due to their low density (2.7 g·cm$^{-3}$ vs. 7.8 g·cm$^{-3}$ for SS) and reasonable yield strength (~290 MPa for AA 2219/T6 vs. ~230 MPa for annealed 304). They are particularly attractive to hydrogen environments because of their extremely low solubility for hydrogen (5.6x10$^{-6}$ mol H$_2$·m$^{-3}$·MPa$^{-0.5}$ for pure Al [2,3] vs. 17 mol H$_2$·m$^{-3}$·MPa$^{-0.5}$ for SS [1,2] at ambient temperature) and hence their relative immunity to embrittlement in dry hydrogen environments [4–7]. However, microstructural defects in Al may trap hydrogen, changing the effective retention and diffusivity of hydrogen in Al alloys significantly [3,4,8–21].

Young and Scully [9] conducted a comprehensive study of hydrogen diffusivity and trapping in pure Al. Using thermal desorption spectroscopy (TDS) on as-received, cold-worked, and annealed Al wire, they derived the binding energies of hydrogen to vacancy (27.3 kJ·mol$^{-1}$) and dislocation (68.6 kJ·mol$^{-1}$) traps. These deep and plentiful traps are likely responsible for large discrepancies in the apparent diffusivity of and solubility for hydrogen in Al reported in the literature. The concentration of trapped hydrogen often exceeds the concentration of soluble hydrogen [3,8].

The precipitates and solid solution of Al alloys may act as additional trap sites. In the 2xxx series of alloys and in the Al-Cu binary alloys considered here, age hardening forms Al-Cu intermetallic precipitates that strengthen the material. Previous work investigating Al-Cu precipitates has determined the temporal evolution of the precipitates phases [22]:

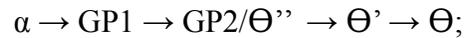
$$\alpha \rightarrow GP1 \rightarrow GP2/\Theta'' \rightarrow \Theta' \rightarrow \Theta;$$

where α is the Al-Cu solid solution, GP1 zones are planes of Cu typically oriented along the {0 0 1} α-Al plane. As Cu continues to leave the solid solution, GP2/ Ө'' zones form as "sandwiches" of two Cu disks in the {0 0 2} layers separated by three Al planes. These metastable precipitates evolve slowly into intermetallic spherical precipitates: transitioning first to the body-centered tetragonal Ө' $Al_2Cu$ phase and ending eventually with the equilibrium incoherent Ө $Al_2Cu$ precipitates that have a tetragonal *C*16 crystal structure. Each of these phases exerts a tensile stress fields and are likely traps for hydrogen isotopes.

Increasing Cu content leads to an increase in the amount of hydrogen retained in solid Al-Cu alloys [23]. Internal friction experiments have placed an upper bound for a binding energy to Cu in solution to 5 kJ·$mol^{-1}$ (0.05 eV) [24]. Tritium autoradiography experiments have shown that in Al-Cu (unlike Al-Si alloys), all of the metastable and stable precipitate phases seem to trap hydrogen [17].

Detailed analysis of trapping energies using TDS is often complicated by multiple phases with similar trapping energies, and because the temperatures accessed during charging and TDS often cause the coarsening and/or dissolution of precipitates. The microstructural features responsible for particular trap binding energies are inferred from relative energies and occupancies based on systematic compositional or thermo-mechanical changes made to the materials. This becomes more and more difficult as the number of different microstructural features that may serve as trap sites increases. Further, typical analysis of TDS presumes a first order desorption process, a condition that may not be met if binding energies are larger than the energies required to cause coarsening or dissolution [25–27]. Other alloys, such as steels, share these challenges and researchers have relied on differential scanning calorimetry (DSC) and X-ray diffraction to further support their inferences [28].

Hydrogen-trapping energies for various phases in Al-Li-Cu-Zr have been determined, but the team that performed the study did not observe significant trapping in binary Al-Cu alloys [25]. Nonetheless, traps have been attributed to not only $Mg_2Si$ ($\beta$), but also $Al_2Cu$ ($\theta$) phases in AA 2024/T3 (Al-4.35 Cu-1.5 Mg (compositions in wt.%; Mn composition not specified, but presumably 0.5-0.6)) [10]. A subsequent study of AA 2024/T351 attributed the higher energy trap observed there instead to Mg-containing S-phase ($Al_2CuMg$) [13]. Neither work determined the binding energy or occupancy of the trap, however.

Due to the discrepancies in trapping measurements to Al-Cu precipitates in the literature, this work uses high pressure (140 MPa) deuterium ($D_2$) charging at elevated temperatures (200-300 °C), variable ramp rate TDS, electrical conductivity, and DSC to study the binding energy and typical occupancy of these precipitate traps.

## 2. Experimental

### 2.1. Materials and Deuterium Charging

[Table 1 HERE]

Nine different castings of Al-$X$ Cu (0.0 at. % < $X$ < 3.5 at. %) were cut into strips ca. 2 cm x 2 cm x 1.34 mm. The greatest impurities were determined to be Si, Fe, and Zn; all are less than $7 \times 10^{-3}$ at. %. Compositions of these are given in Table 1. Deuterium was charged into the sample in 130 MPa $D_2$ for either 68 days at 200 °C or 30 days at 300 °C. The expected bulk concentration (mol D/mol Al) in pure Al would be $4.45 \times 10^{-7}$ and $2.59 \times 10^{-6}$ for these respective methods [2,3]. The extended treatment at elevated temperatures is expected to create Θ' and Θ precipitates at both aging temperatures (GP zones are only observed at lower temperature aging and for aging shorter than a day) [22].

Upon the completion of the charging, the furnace is air cooled to ambient temperature and the samples are immediately placed in a freezer of -54 °C where D diffusion in the sample is very slow ($1.7 \times 10^{-12}$ m²/s [2,9]). Deuterium is an isotope of hydrogen with a single neutron; it was chosen for this study because it is not naturally abundant in the earth (only 0.0115% of natural hydrogen is deuterium) and environmental deuterium in the vacuum system is well below the detection limit of the residual gas analyzers used in this experiment. Deuterium is a good model for hydrogen interaction in the material, having the same solubility and trapping and a diffusivity, $D$, that is known to scale as $D_D = D_H \sqrt{\frac{m_H}{m_D}}$, where $m$ is the mass of the respective isotope and the subscripts $D$ and $H$ refer to deuterium and hydrogen, respectively.

## 2.2. Electrical Conductivity

The electrical conductivity ($\sigma$) measurements were taken at ambient temperature with an SigmaCheck Eddy Current conductivity meter manufactured by ETHER NDE. Four separate measurement frequencies (60, 120, 240 and 480 Hz) were used on every specimen and the meter was calibrated independently at each measurement frequency using pure Al and pure Cu. Changing the frequency changes the depth of the measurement and is done to ensure the through-thickness homogeneity. Uncharged specimens with less than 1.3 at. % Cu were solutionized at 550 °C for 72 hrs to establish the dependence of electrical conductivity on the Cu in solution. The maximum solid solubility for Cu in Al is 2.4 at. % [29], so these samples are assumed to be fully solutionized.

Electrical resistivity, $\rho$, is the inverse of electrical conductivity, $\rho=1/\sigma$, and increases proportionally to solute content, $C$. As precipitates nucleate and grow, $C$ decreases, decreasing $\rho$ and increasing $\sigma$. Charged specimens were measured to determine these electrical conductivity changes due to precipitate evolution during charging and/or the D present in the materials. Finally, samples were tested after various segments of thermal ramps to determine the solute/precipitate temporal evolution.

## 2.3. Differential Scanning Calorimetry

In this study, a STA 449 F3 Jupiter manufactured by Netzsch was used for DSC measurements. By calibrating with both an inert reference (empty crucible) and a sapphire sample of known heat capacity, the data for the heat capacity of Al-Cu samples of interest are also obtained. In this study, uncharged Al-Cu samples were used to identify the temperature range where the precipitates dissolve.

## 2.4. Thermal Desorption Spectroscopy

The mass and electrical conductivity of the samples were measured before and after the thermal desorption spectroscopy (TDS) experiments at room temperature. A scale accurate on the order of micrograms manufactured by Mettler Toledo measured the masses of the samples. On average, the total transfer time between removing the sample from the freezer to starting the TDS system was 15 min. Much of this time was spent in pumping down the TDS system at ambient temperature.

[Figure 1 HERE]

The system used in these experiments is shown schematically in Figure 1. Constant heating rate tests from 0.5 °C/min to 10 °C/min were performed. To start, a sample was placed in the loading chamber in the quartz furnace at room temperature. After sealing the system, the chamber was then pumped down to around 12 mtorr before the barrier between the ultra high vacuum chamber (ca. 1 x $10^{-9}$ torr) was opened. The experiment began when the chamber achieved a pressure of 1x$10^{-6}$ torr. A residual gas analyzer (RGA) was used to monitor the partial pressure of D (from both mass 3 (HD) and mass 4 ($D_2$)) outgassing from the sample. The signal from these peaks was large enough that the Faraday cup was able to detect these without the use of a channel electron multiplier. The furnace temperature was measured by a thermocouple in the quartz thimble that the tantalum sample holder was attached onto. These Al-Cu samples were heated to a maximum temperature not exceeding 550 °C to avoid melting.

In addition to the ramped TDS experiments, interrupted TDS experiments were also conducted. The furnace was programmed to bring the temperature of the sample to the intended start temperature held isothermally for 5 min to insure temperature stabilization then is heated at a predetermined constant rate to the incremented stop temperature and also held isothermally before cooling to room temperature. The temperature increment was 100 °C – so for these samples, the TDS was interrupted 3 times (at 200 °C, 300 °C, and 400 °C). In between each of these increments, the mass and electrical conductivity of the sample were measured.

A multi-peak Polanyi-Wigner fit implemented in GNU Octave [30] to the TDS data determines trapping peaks. The Polanyi-Wigner equation for desorption rate, R(t), is [31]:

$$R(t) = \upsilon N(t)^\beta \exp(-E_d/k_B T); \qquad (1)$$

Where $\upsilon$ is the attempt frequency, $N(t)$ is the number density of reactants on the surface, $\beta$ is the order of desorption, $E_d$ is the effective activation energy for the dominant recombination and desorption process, and $T = T(t)$ is the sample temperature. While a multi-variable fit to this equation may provide a reasonable estimate of the trapping energy from a single desorption rate, changing the desorption rate may reduce the free parameters and reduce the uncertainty of the derived trapping energy. When the sample is subject to an increasing temperature ramp, it will desorb D from the trap with the lowest to highest binding energy. If all conditions are maintained and the heating rate is adjusted, the peaks will shift; this shift can be plotted on a Kissinger plot to determine the binding energy [32]. As the heating rate increases, desorption peaks shift to higher temperatures. It is possible to find $E_d$ by calculating the slope of the best fit line through the data in a semi-log plot:

$$E_d = R \frac{d \ln \frac{\Phi}{T_m^2}}{d \frac{1}{T_m}}; \qquad (2)$$

Where $R$ is the ideal gas constant, $T_m$ is the peak temperature, and $\Phi$ is the thermal ramp rate.

## 3. Results

### 3.1. Electrical Conductivity

[Figure 2 HERE]

Figure 2 shows the electrical resistivity of solutionized samples ($C$<1.29 at. % Cu). A linear fit to this data shows that:

$$\rho = (7.6 \pm 0.1) \times C + (26.60 \pm 0.08); \quad (3)$$

where $\rho$ is given in nΩm and $C$ is given in at. % Cu. The intercept of 26.60 ±0.08 nΩm is equivalent to an electrical conductivity of 37.6±0.1 and is in good agreement with the values reported for pure Al at ambient temperature (36.6-37.7 MS/m [29,33,34]). The slope of (7.6±0.1) nΩm/at. % Cu is also in reasonable agreement to literature of dilute Al-Cu alloys [34,35].

[Figure 3 HERE]

Electrical conductivity is also used to track the temporal evolution of the material during charging runs. As seen in Figure 3, the thermal aging at 200 °C and 300 °C increases the conductivity by 4 and 2 mS/m respectively for the sample with the lowest amount of copper. This is likely due to precipitate nucleation and growth that deplete the solute Cu concentration. There is a larger change after aging at 200 °C than at 300 °C, possibly due to the extended time at the lower temperature. Further, D in solution and traps does not change the electrical conductivity substantially, as samples subjected to the same thermal treatment in air recorded the same changes in electrical conductivity. This is likely due to D residing in interstitial sites (rather than the substitutional sites where Cu resides) and due to the much lower concentrations of D than Cu. Throughout the rest of the article, we assume the conductivity is wholly determined by the amount of Cu in solid solution.

[Figure 4 HERE]

The incremented TDS experiments (Figure 4) show less than 1 mS/m decrease in electrical conductivity between ambient temperature and 300 °C, but show around 7-8 mS/m decrease between 300 °C and 500 °C. This is due to the dissolution of precipitates, which results in the increase of Cu in solution.

## 3.2. Differential Scanning Calorimetry

[Figure 5 HERE]

Using DSC, the evolution of the precipitates can be observed. The heat capacity results (Figure 5) show relatively no microstructural changes until ca. 350 °C, when the Cu precipitates begin to dissolve because the precipitate dissolution increases the heat capacity. This supports the notion that the as-charged microstructures are likely to contain a high fraction of stable ϴ precipitates and virtually no GP zones that would make themselves evident from changes at lower temperatures [36,37]. This is as we expect from the very long charging times.

## 3.3. Thermal Desorption Spectroscopy

[Figure 6 HERE]

[Table 2 HERE]

In Figure 6, three distinct trap states are seen from the peak fit's good agreement with the desorption data. Lattice and dislocation traps are inferred from their good agreement with

Ref. [9]. In other spectra measured at higher ramp rates, a peak that is in good agreement with the vacancy peak is observed. Here, though, the expected position of the vacancy peak is above the maximum temperature of our desorption experiments that is set to prevent sample melting. The precipitate trap is inferred here because it is unique to this study and the peak area increased with increasing Cu content. At lower ramp rates (< 0.83 °C/min), a small peak appears between the peaks attributed to vacancies and dislocations that is not observed at higher ramp rates. Due to the high temperatures (>350°C) that it is observed at, it may be associated with precipitate dissolution. This would likely not be a first-order process and we observed this extra peak in only a limited number of experiments, so we make no effort to deduce its origin further.

Our analysis of the other peaks is consistent with first order desorption from dilute concentrations of traps. Making this assumption allows us to estimate the amount of D found in each trap site by integrating the TDS peaks. This is reported in Table 2. The concentration of $D_2$ desorbed from the α-Al lattice is comparable to the expected $D_2$ solubility in pure Al [3] and there is almost as much $D_2$ in each of the two trapping peaks.

[Figure 7 HERE]

As seen from Figure 7, the general shape of spectra does not change (consistent with our assumption of a dilute concentration of first order traps). Each peak position shifts with a change in thermal ramp rate and can be followed on a Kissinger plot.

## 4. Discussion

### 4.1. Trapping in Al-Cu vs. Pure Al and a Comparison of Charging Techniques

[Figure 8 HERE]

Figure 8 displays both the desorption results at 10 °C/min from pure Al reported in Ref. [9] and from Al-2.6 at. % Cu of this study with trap states labeled from low temperature to high. The Al-Cu sample here was charged with D at 200 °C for 68 days, whereas the pure Al sample in Ref. [9] was charged with H at ambient temperatures using an electrolytic cell. Despite these differences in charging methodology, the three peak positions present in both samples are in good agreement. The experimental procedure for placing the hydrogen in the trap sites may slightly differ, but both result in hydrogen at trap sites. The mechanics of hydrogen desorption are the same using TDS; however, there is an addition peak in the fit for the Al-Cu sample at around 325 °C. This peak position appears consistent throughout the ramped TDS data for all runs in this work and the size increases with increasing Cu content. This is consistent with literature reports of hydrogen retention in Al-Cu alloys [23]. The good agreement between the other peaks observed in Ref. [9] provides additional evidence that there is, indeed, an additional trap site present in the Al-Cu alloy.

[Figure 9 HERE]

The results of the various Al-Cu samples charged at 200 °C for 68 days and desorbed at 2.22 °C/min are presented in Figure 9. The total amount of D increases with Cu content and we identify this increase occurs predominantly in the peak we attribute to precipitates, as well as the dislocation peak. Precipitates can increase the dislocation density not only through more misfit dislocations around semi-coherent precipitates, but also by generating them due to the differences in the coefficients for thermal expansion between the α-Al matrix and the precipitates.

[Figure 10 HERE]

The charging conditions of the samples do not greatly alter the peak position of the traps. Instead, it affects the amount of D trapped. As seen in Figure 10, the higher temperature charging conditions resulted in more D trapped as noted by the larger area under the curve, consistent with an expected increase in D solubility in Al and in trap occupancy (and, in the case of, e.g. thermally-created dislocations, an increase in the number density of traps).

## 4.2 Trap Sites in Al-Cu Alloys

[ HERE]

[Table 3 HERE]

The desorption energy is calculated using Eq. (2) on Figure 11 and is presented in Table 3. The trap binding energy, $E_b$, is estimated, as in Ref. [9], by subtracting the activation energy for diffusion of H through the Al matrix (found there to be 16.2±1.5 kJ/mol). There is good agreement of peak positions in individual runs attributed to the Al lattice (1$^{st}$ peak), dislocations (3$^{rd}$ peak), and vacancies (5$^{th}$ peak listed in Table 3, though some ramp rates didn't allow us to fit the peak labeled as '4'). Additionally, we find our fitted values of $E_b$ for dislocations and vacancies are in excellent agreement with Ref. [9], despite the different alloy, charging conditions, and total content of hydrogen isotopes. Likely because of the smaller total concentration of hydrogen isotopes in this study than that one (and therefore the higher relative background below ca. 200 °C), the lattice peak cannot be fit in the slower ramp rate tests here, so we do not consider the two ramp rates here where we did fit it clearly to be adequate at estimating $E_b$ associated with lattice diffusion.

The 2nd desorption peak in Table 3 was observed in every desorption experiment in this study, but does not appear in studies of pure Al. It most likely originates from Cu-containing precipitates, most notably the θ precipitates that are predominant after aging at 200-300 °C for an extended period of time [22]. The binding energy is much greater than the upper bound for Cu in solid solution (5 kJ·mol$^{-1}$) [24]. Further, the amount of D in the trap increases with Cu content. While the volume fraction of precipitates increases with Cu content, the amount of Cu in solution does not (per our conductivity measurements). The peak was not reported in Al-Cu binaries with more modest aging treatments [25], which provides additional support for the conjecture that the trap is due to θ precipitates that appear only after extended aging. The peak seems to be absent from the data for AA 2024 [10,13], perhaps due to the lower volume fraction of θ Al-Cu precipitates at the expense of Mg-containing S-phase and due to not being aged for as long. Some trap site due to these Al-Cu precipitates is consistent with tritium autoradiography experiments of Al-Cu alloys [17]. The modest binding energy suggests that θ would not affect the effective diffusivity of hydrogen isotopes much in the relatively dilute commercial alloys.

We do not identify the 4th desorption peak in Table 3. It is likely somehow associated with Cu, given that it is absent in the desorption spectra of pure Al. It is also absent at high thermal ramp rates in this study. This may be associated with GP zones that could be formed by extended times at moderate temperatures, but this seems somewhat unlikely from our electrical conductivity of interrupted TDS and our DSC measurements. The high temperature of the peak suggests that it may be correlated with precipitate dissolution that would also occur more over the longer runs at the lower heating rates. If associated with dissolution, the peak does not correspond to a simple 1st order desorption process and complicates binding energy analysis. A similar apparent trapping energy is found in Cu-containing AA 2024 in both Ref. [10] (where the authors attribute it to Mg, that isn't present in our study) and Ref. [13], where the authors attribute it to Mg-containing S-phase (Al$_2$CuMg). Neither work determines the binding energy of each trap site. However, we estimate it with Eq. (2) to be 46 kJ/mol, which is consistent with the value calculated for the 4th peak of this study.

Past authors who observed trapping in Al-Cu precipitates did not estimate a trapping energy associated with precipitate phases, but we have attempted to understand our data in the context of these past experiments. We have analyzed the data in Ref. [10] by taking peak maxima and assuming the authors used the ramp rate of between 5-6 °C/min and present this in Fig. 11 in the form of error bars corresponding to the range of the heating rate. Four of the peaks in that paper do seem to match well with peaks, but the interpretation of these differ from the discussion there, so this warrants a more detailed analysis. There is a low energy peak in that data labeled 'T1' at ca. 100 °C reversible trap observed there that is not observed in the most other studies, but may be present in Ref. [13]. The authors of Ref. [10] dismiss the possibility of it being adsorbed hydrogen because they expected no peak associated with adsorbed hydrogen, but the very low energy would be consistent with that or some other reversible process. The second peak, 'T2', at ca. 250 °C is more consistent with the Al lattice peak of this study and in Ref. [9] and this seems to be a more likely mechanism than trapping by β Mg$_2$Si precipitates, as the authors of Ref. [10] speculate. This discrepancy is likely due to the mis-assignment of T1. The next three peaks (unlabeled peaks at ca. 350 °C and 425 °C and 'T3' at ca. 450 °C) seem to correspond well with what we've labeled as dislocations, an unidentified peak associated with Cu, and vacancies, respectively. There is a final large peak in that work that does not appear here, 'T4' at ca. 550 °C that the authors attribute to the dissolution of θ Al$_2$Cu precipitates. We think it may instead be due to the Mg because (i) we do not see the peak in our Mg-free alloys; (ii) we see dissolution θ precipitates at lower temperatures; (iii) in other studies, we've seen that Mg has a high affinity for H in Al; and (iii) in those studies, we've observed Mg evaporation near this temperature. That Mg plays some role in this peak is also reinforced by Ref. [13], that attributes the T4 peak to the Mg-containing S-phase precipitates.

## 5. Conclusion

This study of Al-$X$ Cu (0.0 at. % < $X$ < 3.5 at. %) that has been charged with 130 MPa D$_2$ for either 68 days at 200 °C or 30 days at 300 °C and a thorough exploration of the existing literature for hydrogen in Al and Cu-containing Al alloys allows us to reach the following conclusions:

- As in pure Al, dislocations and vacancies act as traps for D in binary Al-Cu alloys with similar trapping energies.

- There is further trapping associated with increasing Cu content that we attribute to Ɵ Al$_2$Cu precipitates with a binding energy of 17.6 ± 2.5 kJ·mol$^{-1}$ (0.19 ± 0.03 eV).

- For Al-2.6 at. % Cu (a Cu composition comparable to that in AA2219) charged at 200 °C with 130 MPa D$_2$ for 68 days, there is ca. there is 3.15x10$^{-7}$ mol D bound to this precipitate trap per mol of Al, accounting for a third of the D in the charged sample.

- There is an additional apparent high binding energy (50.0 ± 8.6 kJ·mol$^{-1}$ if first order desorption is assumed) trap that is absent from studies of pure Al, but appears in other literature reports for Cu-containing alloys. This apparent trap contains relatively little D and it may be associated with precipitate dissolution, based on the high temperatures that it is found at.

# Acknowledgements

Drs. D.K. Ward, C. San Marchi, B.P. Somerday, and D.K. Balch are thanked for helpful discussions. The authors gratefully acknowledge support from the Sandia National Laboratories Laboratory Directed Research and Development (LDRD) Program. Sandia is a multi-program laboratory managed and operated by Sandia Corporation, a wholly owned subsidiary of Lockheed Martin Corporation, for the US Department of Energy's National Nuclear Security Administration under Contract No. DE-AC04-94AL85000.

# TABLES

## List of tables

Table 1: Compositions of Al-Cu alloys studied

| Alloy Name | Cu [at. %] | Si [at. %] x $10^{-4}$ | Fe [at. %] x $10^{-4}$ | Zn [at. %] x $10^{-4}$ |
|---|---|---|---|---|
| Al-0.0 Cu | 0.00068 | 23. | 20. | 2.0 |
| Al-0.1 Cu | 0.0816 | 29. | 13. | 19. |
| Al-0.2 Cu | 0.209 | 42. | 14. | 25. |
| Al-0.4 Cu | 0.414 | 48. | 15. | 37. |
| Al-0.8 Cu | 0.837 | 31. | 16. | 17. |
| Al-1.3 Cu | 1.28 | 37. | 17. | 19. |
| Al-1.7 Cu | 1.70 | 39. | 17. | 25. |
| Al-2.6 Cu | 2.61 | 50. | 17. | 5.1 |
| Al-3.4 Cu | 3.44 | 54. | 17. | 64 |

Table 2: Peak positions and amount of $D_2$ desorbed from Al-2.6 at. % Cu charged at 200 °C and desorbed at 5 °C/min.

| Trap Site[*] | Peak Position [°C] | Concentration of D [mols D/mols Al] |
|---|---|---|
| Lattice | 234 | 3.80x$10^{-7}$ |
| Precipitates | 329 | 3.15x$10^{-7}$ |
| Dislocations | 383 | 3.00x$10^{-7}$ |
| Total | -- | 9.95x$10^{-7}$ |
| Expected D in Lattice [3] | -- | 4.61x$10^{-7}$ |

[*] Lattice and dislocation traps are inferred from their good agreement with Ref. [9]. The vacancy peak found there occurs above the maximum temperature of our desorption experiments, which is set to prevent sample melting. The precipitate trap is inferred here because it is unique to this study and the peak area increased with increasing Cu content.

Table 3: Summary of trapping data determined from TDS

| Desorption Peak | Trap Site | Pure Al Desorption Energy, $E_d$ [kJ/mol] [9] | Al-Cu Desorption Energy, $E_d$ [kJ/mol] | Binding Energy, $E_b$ [kJ/mol] |
|---|---|---|---|---|
| 1 | Lattice | 15.3 ± 4.8 | --[*] | -- |
| 2 | Precipitates | -- | 33.8 ± 2.5 | 17.6 |
| 3 | Dislocations | 43.5 ± 17.5 | 40.5 ± 6.0 | 24.3 |
| 4 | ?[**] | -- | 50.0 ± 8.6 | 33.8 |
| 5 | Vacancies | 84.8 ± 32.2 | 68.8 ± 6.6 | 52.6 |

[*] The lattice desorption energy calculated here is 36±12 kJ/mol, but the fitting error does not reflect the high uncertainty due to a poor ability to fit a peak at lower desorption temperatures.

[**] A modest peak that is seemingly between the peak attributed to dislocations and vacancies is visible only at low heating rates. Due to the high temperatures involved, it may be correlated with precipitate dissolution.

# FIGURES

## List of Figures

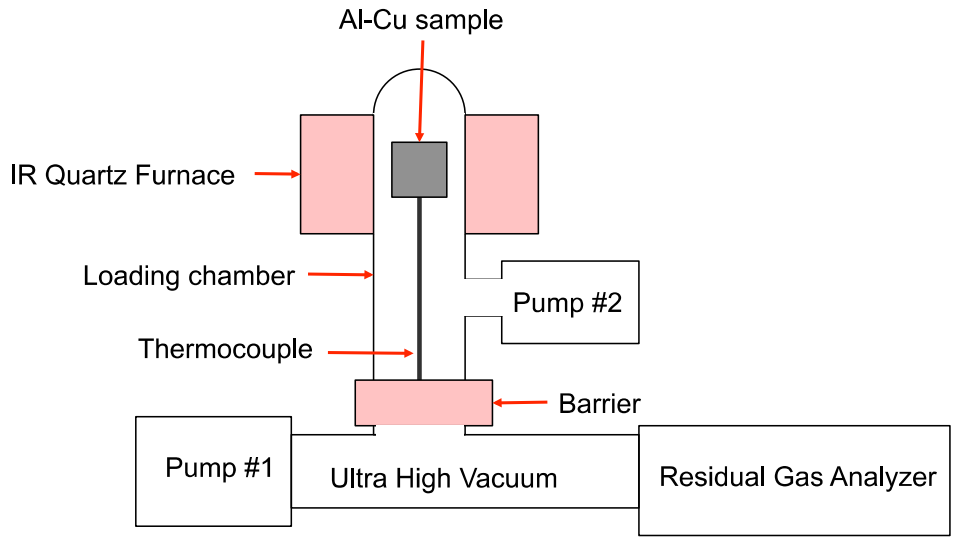

Figure 1: Schematic of the TDS system used. The sample is heated in an infrared quartz furnace and rests in a sample holder mounted to a thermocouple to measure the temperature.

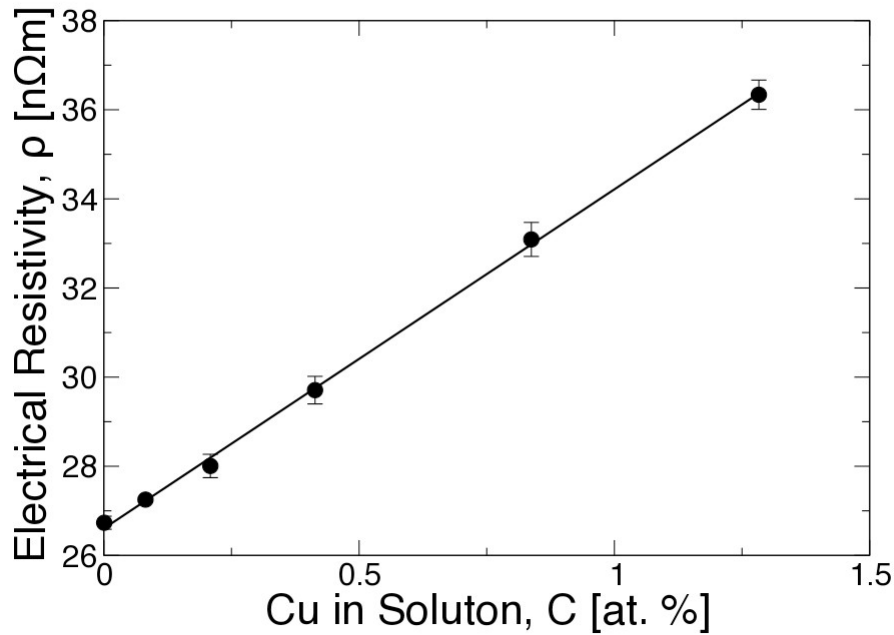

Figure 2: Electrical resistivity, $\rho$, of solutionized Al-$C$ Cu ($C$ < 1.3 at. %). The linear relationship $\rho = (7.6 \pm 0.1) \times C + (26.60 \pm 0.08)$ allows the determination of the amount of Cu in solution in charged or partially desorbed samples.

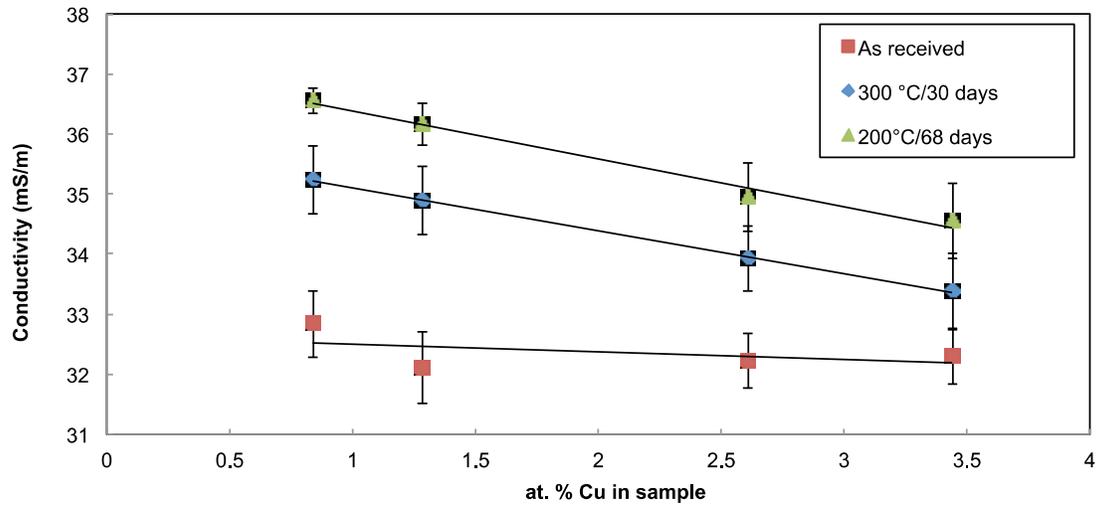

Figure 3: A plot showing the conductivity change of Al-*X* Cu (0.8 at. % < *X* < 3.5 at. %) witness samples after undergoing the 300 °C/30 day or 200 °C/68 days aging in air (the same times and temperatures as in D-charging). The increase in conductivity due to charging indicates that there was precipitate nucleation and growth under the charging conditions imposed. There is more Cu in precipitates in samples that were charged for a longer period of time.

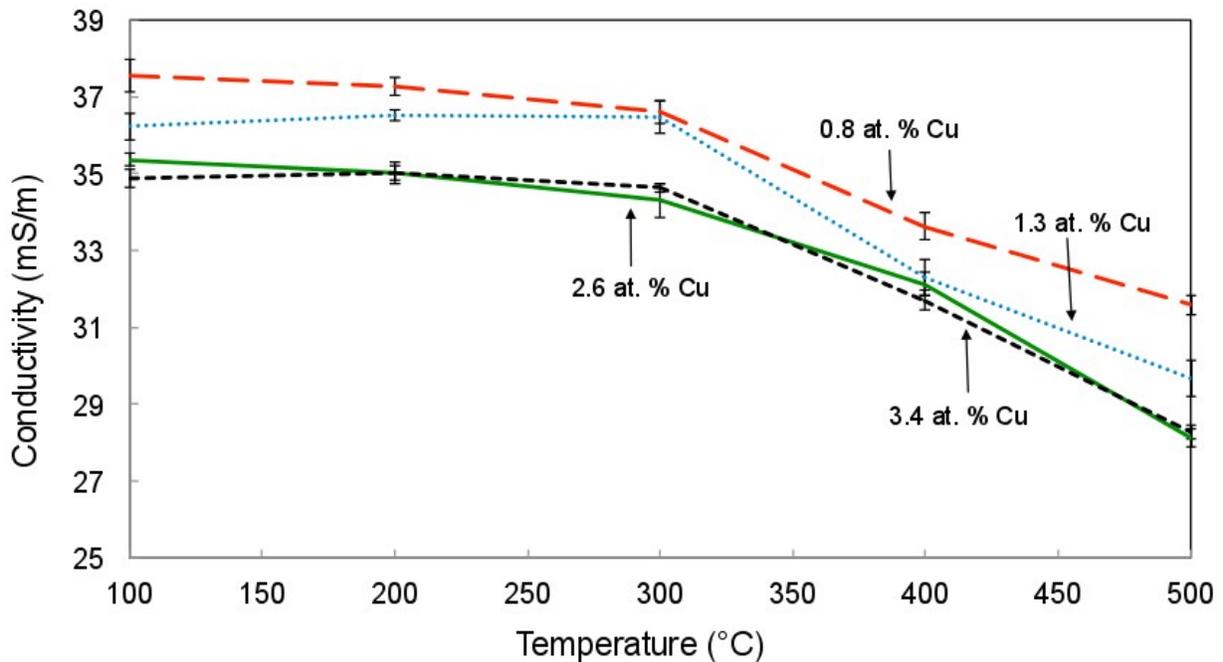

Figure 4: A plot showing the conductivity change for samples charged at 200 °C during interrupted TDS experiments. The electrical conductivity decreases when the sample is raised above 300 °C due to the dissolution of precipitates.

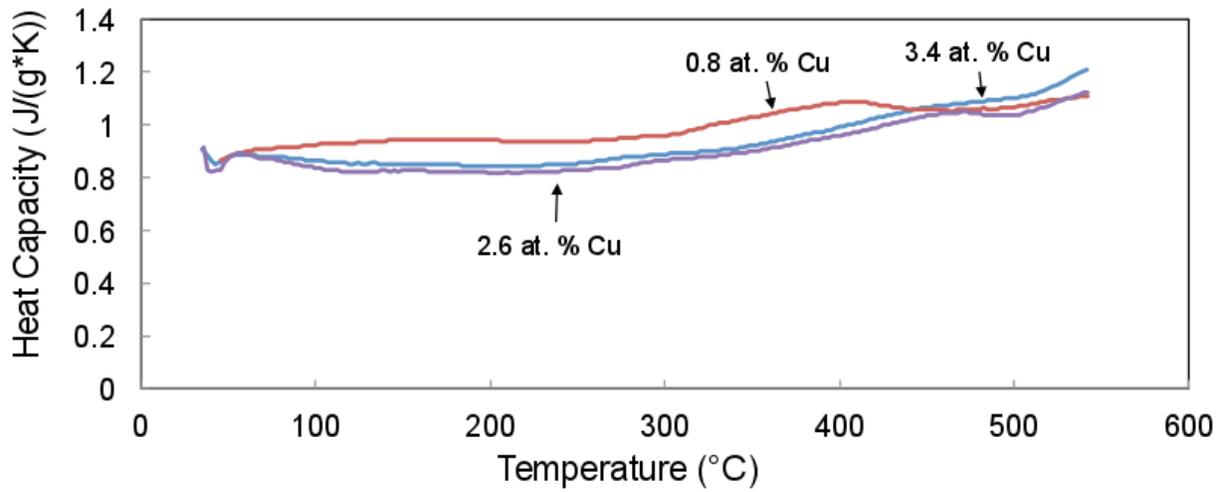

Figure 5: Heat capacity of Al-*X* Cu (0.84 at. % ≤ *X* ≤ 3.4 at. %). There is an increase in heat capacity 300 °C to 500 °C because of the dissolution of precipitates.

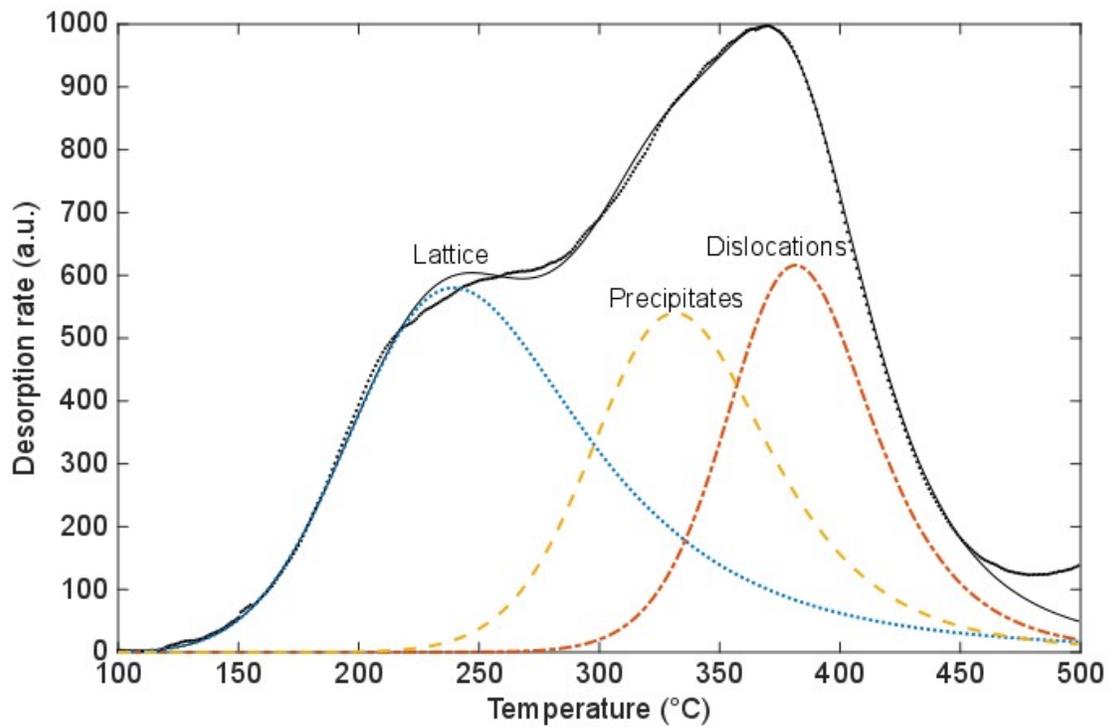

Figure 6: Example of Polanyi-Wigner peak fit of desorption spectra from Al-3.4 at. % Cu charged at 200 °C for 68 days then heated at 5 °C/min. Each peak indicates a trap site and integrating each peak can give the amount of deuterium in each trap site.

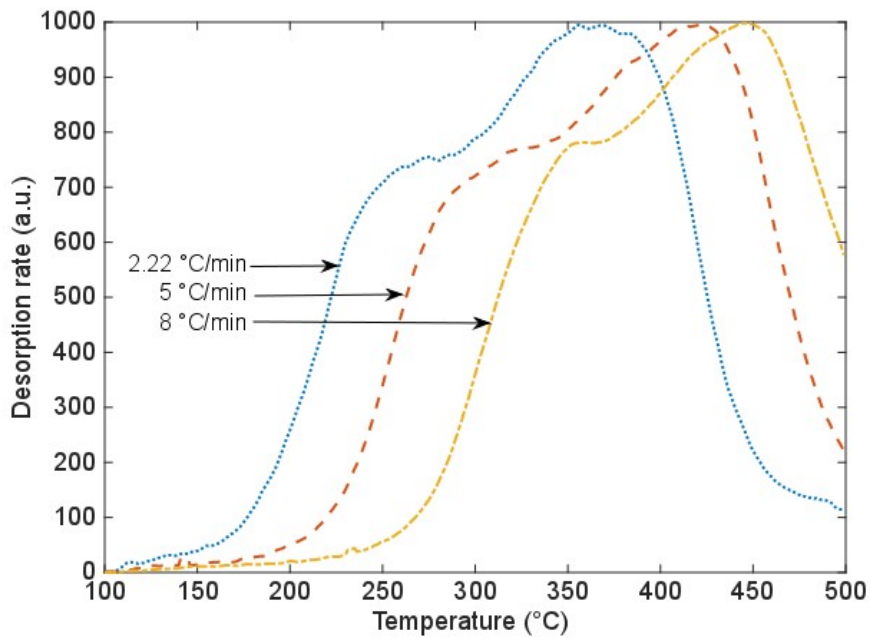

Figure 7: Overlaid spectra of Al-2.6 at. % Cu charged at 200 °C for 68 days and desorbed at various heating rates. The peaks shifts to a higher temperature range due to increasing heating rate. This peak shift can be used to calculate the binding energy of each trap site.

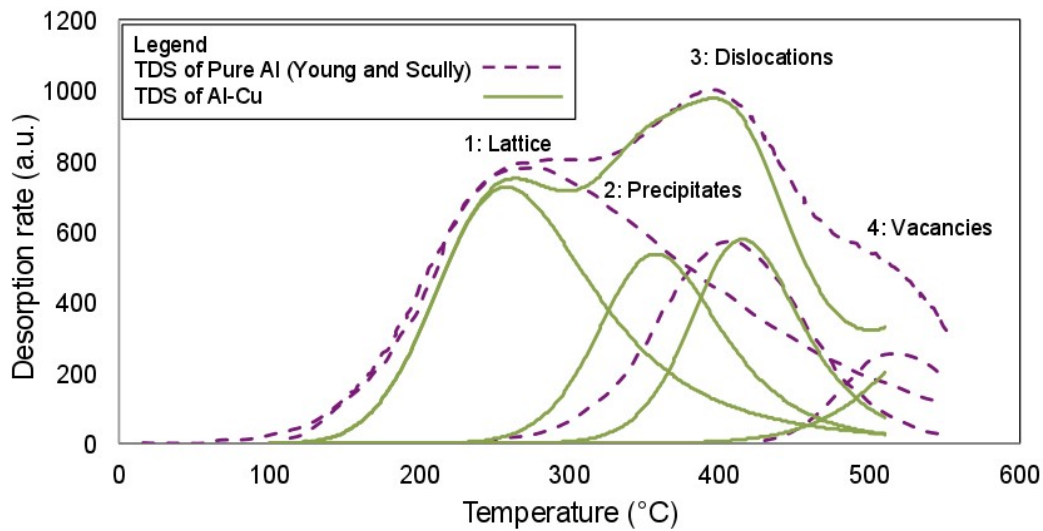

Figure 8: Overlaid desorption spectra of pure Al from Young and Scully [9] and Al-2.6 at. % Cu charged at 200 °C for 68 days, both desorbed at 10 °C/min. The peak fits and determined trap locations are also included. Since pure Al has no precipitate microstructure, the appearance of an additional peak in the Al-Cu spectra suggests that it is a precipitate trap site.

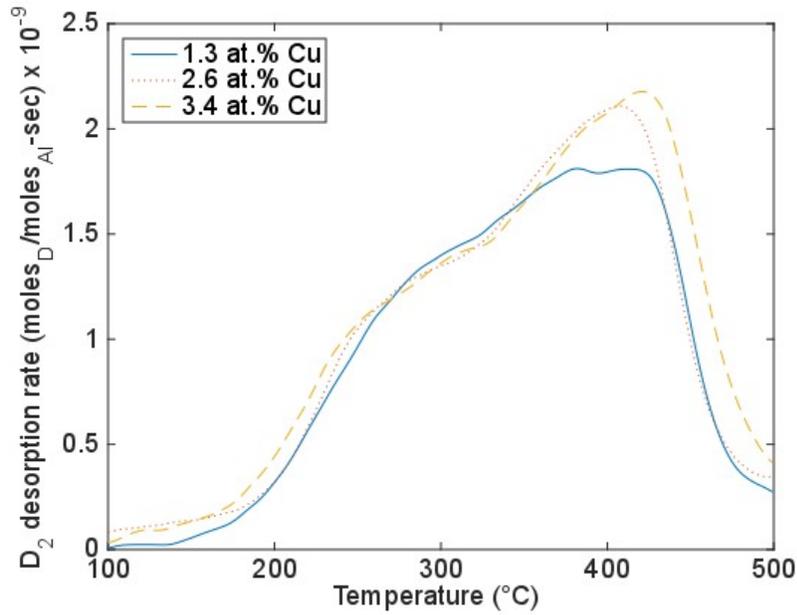

Figure 9: Overlaid desorption spectra of Al-Cu samples of varying Cu content charged at 200 °C for 68 days and desorbed at 2.22 °C /min.

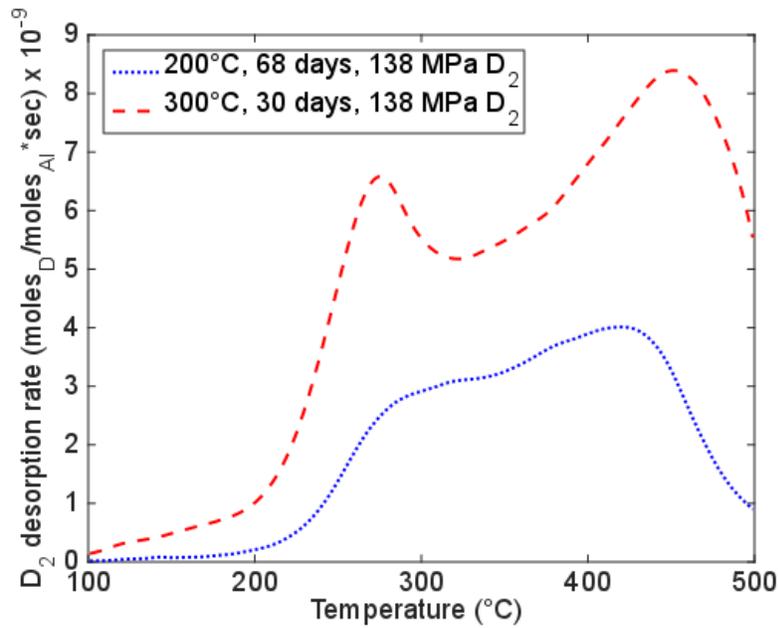

Figure 10: Overlaid desorption spectra tested at 5 °C/min of Al- 2.6 at. % Cu charged at 200 °C and 300 °C. There was more deuterium in the sample charged at 300 °C and the peaks in both desorption spectra are comparable.

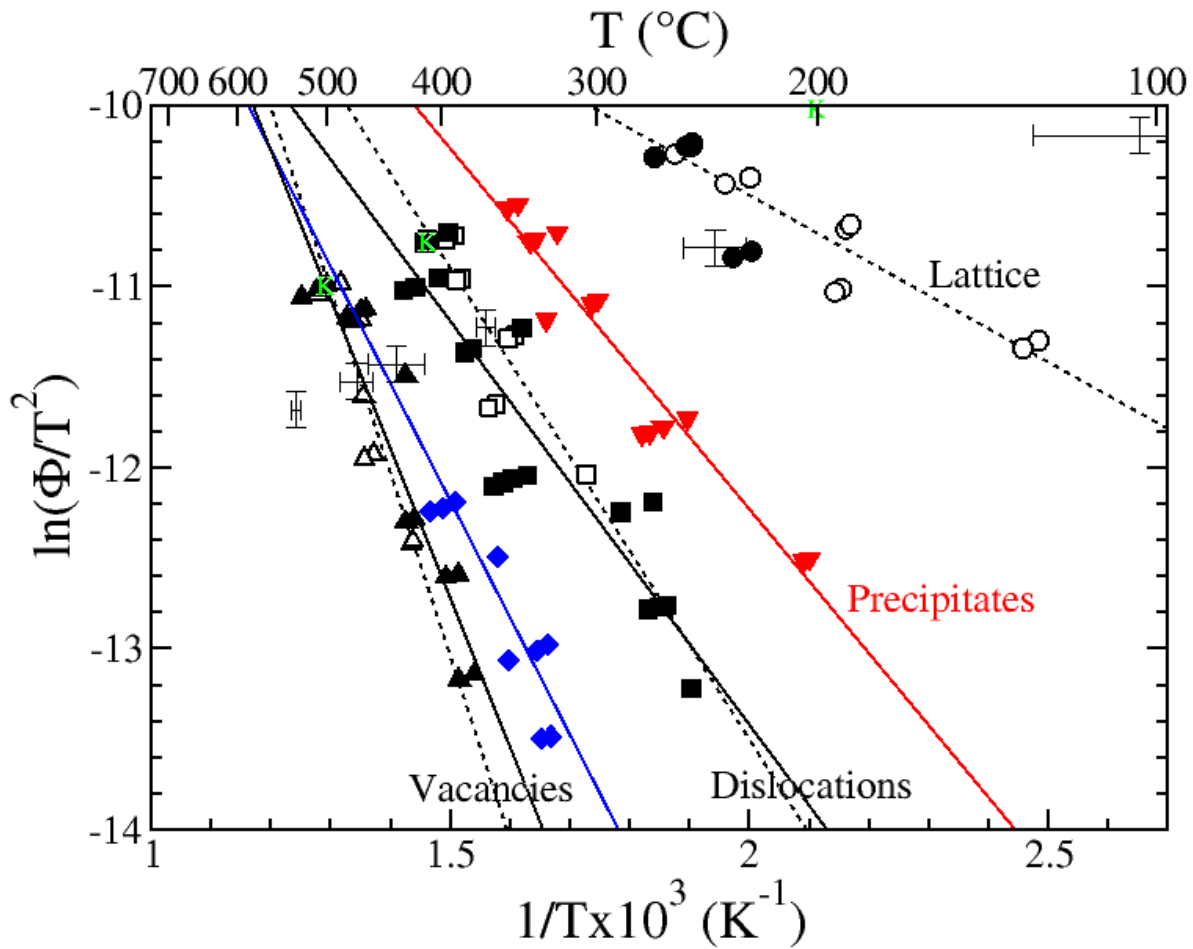

Figure 11: Kissinger plot of the Al (hollow, Ref. [9]), AA 2024 (error bars, extrapolated by authors from Ref. [10]; green 'K' from Ref. [13]), and Al-Cu (solid) desorption peaks for different ramp rates including a 300 °C guideline to indicate the approximate precipitate dissolution temperature. The precipitate trapping peak (red triangles) has not been observed in previous studies. An additional peak (blue diamonds) is observed only in low heating rate tests. Based on the elevated temperatures, it may be related to precipitate dissolution, but we do not attribute a cause for it here.